\documentclass[12pt,a4paper]{article}
\usepackage[dvips]{graphicx}

\catcode`\@=11
\def\v#1{\mathchoice{{\mathord{\mbox{\boldmath$\displaystyle #1$}}}}{{\mathord{\mbox{\boldmath$\textstyle #1$}}}}{{\mathord{\mbox{\boldmath$\scriptstyle #1$}}}}{{\mathord{\mbox{\boldmath$\scriptscriptstyle #1$}}}}}

\def\@makecaption#1#2{\bigskip \footnotesize \setbox1\hbox{{\bf #1.} #2}\ifdim\wd1>350pt\leftskip=30pt \rightskip=30pt {\bf #1.} #2\else\noindent\null\hfill\box1\hfill\null\fi}

\begin{document}

\title{Magnetic monopoles and unusual transport effects in magnetoelectrics}

\author{D.~Khomskii\\[\medskipamount]
\normalsize II~Physikalisches Institut, Universit\"at zu K\"oln,\\
\normalsize Z\"ulpicher Str.~77, 50937 K\"oln, Germany}

\date{}

\maketitle

\noindent
The quest for magnetic monopoles (MM) became a hot topic in
condensed matter physics lately, both theoretically and
experimentally \cite{Moessner,ryzhkin,Bramwell}. Especially exciting is the suggestion of the
appearance of MM on image charges in topological insulators (TI)
\cite{Zhang}. However the experimental setup proposed for their observation
in \cite{Zhang} is rather elaborate,
requiring TI covered by a magnetic layer, using special STM setup, etc.

We propose below that MM should exist in much simpler and
well studied systems -- in solids with the linear magnetoelectric (ME)
effect such as Cr$_2$O$_3$ \cite{Cr2O3}, or in some multiferroics \cite{Khomskii}.\footnote{This
idea was first presented at the Asian-Pacific Workshop on Multiferroics
in Tokyo in January 2011.}
Their existence can lead
to rather striking consequences, such as (magneto)electric Hall
effect, magneto-photovoltaic effect etc.,
which can be observed experimentally.
In addition, in
contrast to the case of TI considered in \cite{Zhang}, in ordinary ME
materials not only MM can accompany the charge, but also more
complicated local magnetic objects can be created, e.g.\ local
toroics, which can also lead to unusual effects in transport and other properties of such systems.

\begin{figure}[ht]
  \centering
  \includegraphics[scale=1.2]{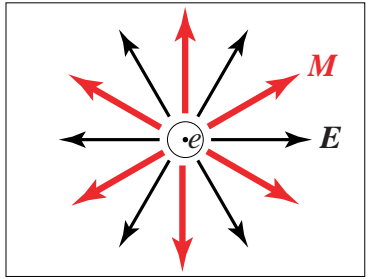}
  \caption{Magnetic monopole (red arrows) created by a charge $e$ in magnetoelectric media with
           the coupling $\alpha\v E\cdot\v B$.}
  \label{fig:1}
\end{figure}

The main idea is actually similar to that proposed in \cite{Zhang}: linear ME coupling of the type
$\alpha\v E\cdot\v B$ can lead to the creation of MM (magnetic ``hedgehog'') by the electric charge.
In contrast to TI where this could happen only on the image charge, in the bulk ME materials the
term of the type $\alpha\v E\cdot\v B$, or even more general $\alpha_{ij}E_iB_j$, exist in the
bulk, with the ME tensor $\alpha_{ij}$ which can have both symmetric and antisymmetric components.
In the simplest case, when $\alpha_{ij} = \alpha\delta_{ij}$, we recover the ME coupling $\alpha\v
E\cdot\v B$ like in \cite{Zhang}.  Then every charge placed in such media, such as an electron or
charge impurity, would create a MM, see fig.~\ref{fig:1}: the electric field
$\v E(\v r) = (e/r^2)(\v r/r)$ would create a radial magnetic polarization $\v M=\alpha\v E$
-- exactly as for a magnetic monopole with magnetic charge $g =
\alpha e$. If an electron would move through such media,
the magnetic monopole would move with it, which can lead to interesting consequences and could
allow to observe such MM\null. The resulting equations describing the dynamics of corresponding
objects will be the same as those for an electric charge, with the substitution
(electric charge~$e$ $\leftrightarrow$ magnetic charge~$g$; electric field~$\v E$ $\leftrightarrow$ magnetic field~$\v H$)
and the linearity of the corresponding equations leads to the well-known
principle of superposition, valid also for the magnetic charges.\footnote{Actually
the radial magnetic field around the charge is created
by the corresponding deviations of ordered spins forming the ME material.
Therefore there will be no real singularity at the centre of~MM\null. 
But the field outside the core will be exactly that of a MM, and all
the dynamic effects will be described by the usual dynamics with the
electric charge $e$ replaced by the magnetic charge~$g$, and the electric
field by~$\v H$.

There will be also no contradiction with the Maxwell equations: as in \cite{Moessner}, the
monopolar-like distribution of~$\v M$ will lead to a similar distribution of magnetic field~$\v
H$, so that the field $\v B = \v H + 4 \pi \v M$ will satisfy the standard Maxwell equation
$\mathop{\rm div}\v B = 0$.}

Some of the resulting consequences are more or less trivial. Thus, one could think
that the motion of a MM in a magnetic field 
would be a special feature proving its
existence.  If we put such material with an extra
electron in an external magnetic field~$\v H$, there will indeed be a force
$\v F=g \v H$ acting on the MM associated with the charge~$e$, with magnetic charge
$g = \alpha e$, i.e.\ the force will be $\v F=\alpha e\v H$. But an external magnetic field
creates in ME media an electric field $\v E = \alpha\v H$, i.e.\ this force will
be $\v F=g\v H=(\alpha e)\v H = e(\alpha\v H) = e\v E$, i.e.\ it is the same force as that of an
electric field acting on a charge. 
In this sense the very motion of an electron in a magnetic field is
not a unique consequence of the appearance of MM.

The question is whether  in this case {\it both forces} will act on the
electron. Thus, if we apply an electric field to such a media, it
will act on the electron by the force $\v F = e\v E$, but this external
field will also create a magnetic field $\v H = \alpha \v E$, and this magnetic field
will also act on the MM, with the force $\v F' = g\v H = e\alpha^2 \v E$.  In effect the effective force
acting on the electron could be combined, $\v F + \v F'$.

\begin{figure}[ht]
  \centering
  \includegraphics[scale=1]{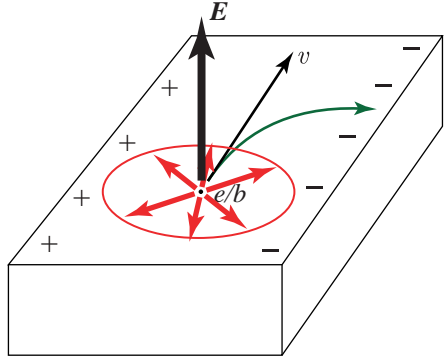}
  \caption{Motion of a magnetic monopole~$g$, accompanying the electric
           charge~$e$, in a perpendicular electric field. The force $g[\v v\times\v E]$ will bend the trajectory
           of this object, leading to an ``electric Hall effect''.}
  \label{fig:2}
\end{figure}

One can also think of other consequences. Thus, similar to the Lorentz force acting on
a charge moving in a magnetic field, $\v F \sim e [\v v \times \v H]$, which leads to the Hall effect,
there will be a similar tangential force acting on a MM moving in an electric field, $\sim\v F
\sim g [\v v \times \v E]$. Consequently, if we put a stripe of such ME material with a current in
a perpendicular {\it electric} field, the MM, together with the electron, would be deviated,
exactly as in the Hall effect, see fig.~\ref{fig:2}.  Such ``electric Hall effect'' should be
observable in ME materials with diagonal ME coupling $\alpha_{ij} = \alpha\delta_{ij}$ and
with~MM\null.
Once again, this may not be surprising:
in such media the electric field $\v E$ perpendicular to the current will also
create a magnetic field parallel to~$\v E$, so that one can also attribute the Hall
effect depicted in fig.~\ref{fig:2} to the ordinary Hall effect in this
effective magnetic field $\v H = \alpha \v E$. Still, this (magneto)electric Hall
effect is an interesting special feature of such ME materials, which
would be very interesting to study.


One can think of some other effects connected with the dichotomy $e$~$\longleftrightarrow$~$g$,
$\v E$~$\longleftrightarrow$~$\v H$. It was proposed in \cite{Nagaosa} to use for TI the
analogy of the classical effect that the electric current creates
a magnetic field around it: the motion of the MM, if present,
would similarly create an electric field around the ``wire'' with the
monopole current, which the authors of \cite{Nagaosa} proposed to observe.
However this could indeed work when the MM is created on the image
charge (the corresponding magnetic field outside the TI, looking like the
field of MM, is actually created by dissipationless currents running
on the surface of the TI). But this would not work in our case: the MM
field around the charge in the bulk of the ME media, fig.~\ref{fig:1}, is in fact
created by a rearrangement of magnetic dipoles in the ME
media, and this field will not exist outside of such sample; in this
sense of course we do not create a real monopole. But, as we argued
above, there will be nontrivial effects connected with the forces and
dynamic of electrons in the bulk of such materials.

Yet another interesting effect may exist in such ME media under
illumination. In ordinary ferroelectrics there exists the well-known
photovoltaic effect \cite{fridkin1}: the electron--hole pairs created under
illumination move in the internal electric field, existing in
ferroelectrics, in opposite directions, thus creating a voltage.
Similarly, in ME media with, for example, a diagonal ME tensor $\alpha_{ij} =
\alpha\delta_{ij}$, there should exist a similar effect caused by the magnetic field (cf.~\cite{fridkin2}):
photoexcited electrons and holes would be accompanied by
the corresponding monopoles with magnetic charges~$\pm g$, which would move in opposite direction in an
external magnetic field (magneto-photovoltaic effect). But here one can also
explain this effect simpler: an external magnetic field will create
an electric field inside the ME material, which will act as the usual
electric field in ferroelectrics, causing the photovoltaic effect.

\begin{figure}[ht]
  \centering
  \includegraphics[scale=1.2]{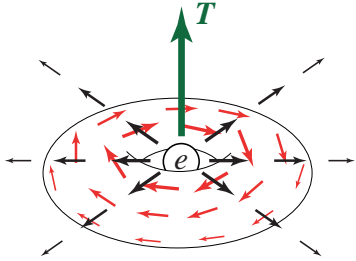}
  \caption{Local toroic (``magnetic donut'') created around
           charge $e$ in magnetoelectric media with antisymmetric magnetoelectric tensor
           and with magnetic moments $\v M(\v r)\sim\v E(\v r)\times\v T$ (close
           to the core $\v M$ should decrease again). Thin radial arrows show
           the electric field, and red arrows show the orientation of magnetic moments around
           the charge~$e$.}
  \label{fig:3}
\end{figure}
\begin{figure}[ht]
  \centering
  \includegraphics[scale=1]{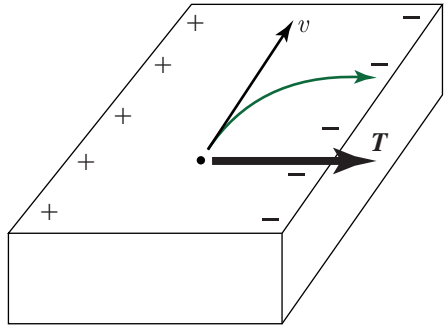}
  \caption{Spontaneous Hall effect in media with
           antisymmetric magnetoelectric tensor $\alpha_{ij}$, or with
           toroidal moment~$\v T$.}
  \label{fig:4}
\end{figure}

The origin of the effects considered above conceptually is similar to
that proposed for TI \cite{Zhang}, although the resulting effects have here a
bulk nature and seem to be more general and easier to observe.
However ME media open yet other possibilities. The ME tensor $\alpha_{ij}$ can have
not only a symmetric, but also an antisymmetric part, which can be
described by the dual vector -- the toroidal, or anapole moment~$\v T$. The
resulting ME coupling can then be written as $\v T \cdot [\v E \times\v B]$. Consequently,
an electric charge placed in such media will create a local magnetic
``toroic'' -- a magnetic vortex with the magnetic moment $\v M(\v r) \sim \v T \times\v E$
(with $\v E = (4\pi e/r^2)(\v r/r)$), see fig.~\ref{fig:3}. The moving electron will carry with
itself such a ``magnetic donut''. To detect it would be less easy than MM:
the toroidal moment couples not to $\v H$, but to ${\rm curl}\,\v H$
\cite{Mostovoy}, $\sim\v T\cdot [\v\nabla \times \v H] = \v T \cdot 4\pi\v j$.
In effect we can have in such media for example a spontaneous Hall
effect like that shown in fig.~\ref{fig:2}, but even without a perpendicular
external electric field: if the toroidal moment is perpendicular to
the current~$\v j$, such current would deviate so as to have a component
parallel to~$\v T$, decreasing the corresponding interaction energy.
This would create a Hall voltage, see fig.~\ref{fig:4}.
Again one can explain this simpler by
saying that the applied voltage and the corresponding electric field $\v E \parallel
\v j$ will create a magnetic field $\v H \sim \v T \times \v E$, and in this magnetic field
there will appear an ordinary Hall effect.  Nevertheless, it is still an interesting
effect, which would be interesting to study experimentally. There should be
probably some other nontrivial effects connected with the presence of
toroidal moments in the system, which are worth exploring.


Summarising, in analogy with  the suggestion of \cite{Zhang}, we
propose that nontrivial local magnetic objects, such as magnetic
monopoles or magnetic toroics, should exist in magnetoelectric media
in the bulk, associated with  electrons or with charged impurities.
Their presence can have nontrivial manifestations,
such as an ``electric Hall effect'' or a magneto-photovoltaic effect --
effects which could be observable experimentally. Thus the ME media
can provide yet another possibility of modelling of magnetic monopoles in solids, more
general and more easily detectable than e.g.\ in frustrated magnets
\cite{Moessner,Bramwell}. One may expect many interesting effects in such systems,
some of which were proposed above. Most of these can be explained both in the picture of
magnetic monopoles and by the more traditional concepts, but the monopole language is 
more transparent and helps to suggest new effects. One could also think of
more special effects for which the monopole picture would be more appropriate.

Note added: Very recently the monopoles in magnetoelectric systems were
also discussed from a different perspective in \cite{Spaldin}, also with the use of
ab~initio calculations for specific materials. I am grateful to N.~A.~Spaldin for 
informing me of this work.

I am grateful to C.Batista and L.Bulaevskii for useful comments.


\end{document}